\documentclass[oldversion]{aa}
\usepackage{graphicx}
\usepackage{natbib}
\usepackage{longtable}

\bibpunct{(}{)}{;}{a}{}{,} 

\begin{document}

\title{Revisiting the connection between magnetic activity, rotation 
period, and convective turnover time for main-sequence stars}

\author{M. Mittag\inst{1}\and J. H. M. M. Schmitt\inst{1}\and K.-P. Schr\"oder\inst{2}}
\institute{Hamburger Sternwarte, Universit\"at Hamburg, Gojenbergsweg
  112, 21029 Hamburg, Germany\\
           \email{mmittag@hs.uni-hamburg.de}
           \and 
           Department of Astronomy, University of Guanajuato, Mexico}
\date{Received \dots; accepted \dots}

\abstract{

{The connection between stellar rotation, stellar activity, and
convective turnover time is revisited with a focus on the sole contribution
of magnetic activity to the Ca II H\&K emission, the  so-called excess flux,
and its dimensionless indicator  R$^{+}_{\rm{HK}}$ in relation to other stellar
parameters and activity indicators.}
{Our study is based on a sample of 169 main-sequence stars with directly
measured Mount Wilson S-indices {\bf and} rotation periods.  The
R$^{+}_{\rm{HK}}$ values are derived from the respective S-indices
and related to the rotation periods in various $B-V$-colour 
intervals.}
{First, we show that stars with vanishing magnetic activity,
i.e. stars whose excess flux index R$^{+}_{\rm{HK}}$ approaches zero,
have a well-defined, colour-dependent rotation period distribution; we also show that this rotation period distribution
applies to large samples of cool stars for which rotation
periods have recently become available.
Second, we use empirical arguments to equate this rotation period distribution
with the global convective turnover time, which is an approach that
allows us to obtain clear relations between the magnetic activity related excess flux index R$^{+}_{\rm{HK}}$, rotation periods, and 
Rossby numbers.  Third, we show that the activity versus Rossby number
relations are very similar in the different activity indicators.}
{As a consequence of our study, we emphasize that
  our Rossby number
  based on the global convective turnover time
approaches but does not exceed unity even for entirely inactive stars.  
Furthermore, the rotation-activity relations might be  universal 
for different activity indicators once the proper scalings 
are used.}
}


\titlerunning{Relation between magnetic activity and rotation}

\maketitle

\section{Introduction}

Magnetic activity of late-type, solar-like  stars is usually discussed in the 
context of the so-called rotation-age-activity paradigm.
The intimate relation between stellar rotation and age was
realized and described in the landmark paper by \citet{Skumanich1972ApJ...171..565S},
who also pointed out the connection between rotation and chromospheric
activity as seen in the cores of the Ca~II H\&K lines.
Early evidence for such a correlation between 
chromospheric activity and stellar age had been
presented by \cite{Wilson1963ApJ...138..832W},
and much later, \cite{Donahue1998ASPC..154.1235D} 
published the first relation describing the connection between 
stellar age and Ca~IIH\&K flux excess R$^{'}_{\rm{HK}}$,
yet it is a common paradigm that rotation is the essential factor
governing all facets of magnetic activity.

As a consequence, numerous studies of activity phenomena of late-type stars 
in various spectral bands 
and their dependence on rotation were carried out starting in the 1980s.   
Another landmark study in this context was the paper by \cite{noyes1984}, who 
demonstrated that at least the chromospheric activity of late-type stars 
could be best understood in terms of the so-called Rossby number, i.e. the
ratio of stellar rotation period to convective turnover time.
The introduction of the Rossby number and hence the convective turnover time 
turned out to be an important step for the understanding of the correlation between
the rotation and activity.
However, the precise meaning of this convective turnover time is 
less clear. The first empirical definitions of the convective turnover time made use
of the Ca II H\&K excess flux \citep{noyes1984}, while in the last decade
the X-ray flux has often been used for an empirical 
definition \citep{2003A&A...397..147P}. 

In this paper we focus on the purely magnetic activity related
fraction of Ca~II~H\&K chromospheric emission (sometimes referred to as 
flux excess) and revisit the correlation between the respective
R$^{+}_{\rm{HK}}$ index and rotation; the R$^{+}_{\rm{HK}}$ index was introduced by \cite{mittag2013A&A549A117M} to
be distinguished from the former R$^{'}_{\rm{HK}}$ index, which includes 
photospheric and basal contributions. 
By explicitly separating the photospheric and basal
flux contributions to the chromospheric Ca II H\&K emission and separating out
the magnetic activity related component, we can draw a clearer
picture of how activity relates to rotation and convective turnover time.
We specifically define a new, empirical convective turnover time, which
we then use to calculate the Rossby number of given star. With this Rossby number, the
correlation between rotation and activity is revisited. 
When comparing main-sequence stars with different masses and hence
different $B-V$-colours, we must consider the $B-V$ dependence in the relation
between the stellar rotation and activity.  Our paper is organized as follows:
In Sec.~\ref{sec_sample} we define the sample of stars to be studied,
in Sec.~\ref{sec_rhkplus} we show how the magnetic activity related part of the
observed chromospheric emission can be extracted, and in Sec.~\ref{sec_convover}
we introduce theoretical and empirical convective turnover times and compare
the two concepts.  We proceed to show in Sec.~\ref{upper_envel_test}
that there is a well-defined
upper envelope period distribution for very large samples of late-type
stars with measured rotation periods and that this period envelope agrees
very well with theoretical convective turnover times.  Using these
convective turnover times, we can compute Rossby numbers and relate in
Sec.~\ref{sec_multi}
activity measurements in various chromospheric and coronal activity indicators
to each other. 

\section{Stellar sample and observational data}
\label{sec_sample}
For our analysis we used 169 stars selected from a variety of 
catalogues to obtain a sample of solar-like main-sequence stars
with measured rotation periods and chromospheric emission.  More specifically, 
our main selection criteria are, first, that the $B - V$ colour index 
of the star be in the range 0.44~$<$~B~-~V~$<$~1.6, and both the rotation period and the
S$_{\rm{MWO}}$ \citep{Vaughan1978PASP90267V} must have been directly measured.
To assess whether a star is a main-sequence star, its location in the
Hertzsprung-Russell diagram is estimated using the visual magnitude and the parallax from the Hipparcos
catalogue \citep{HIPPARCOS1997ESA}. 
As a threshold to define an object 
as a subgiant or main-sequence star, we used the mean absolute visual magnitude
difference between the subgiant and main-sequence stars 
(which depends on the $B-V$-colour). 

As far as  S$_{\rm{MWO}}$ values are concerned, we prefer the 
mean S$_{\rm{MWO}}$ values as listed in \cite{b95} whenever 
available, assuming a general error of 0.001. These values are the most precise
published average S$_{\rm{MWO}}$ values. If a star is not listed in \cite{b95},
we extracted different S$_{\rm{MWO}}$ values from different catalogues and average
these values. For the errors of these averaged S$_{\rm{MWO}}$ values, we used the standard 
error of the mean. 
For the objects HD36705, HD45081 and HD285690 only one 
measured S-index was found, therefore for these objects a general error of 0.02 
was assumed \citep{mittag2013A&A549A117M}.

In Tab.~\ref{data_tab} we list the
stars selected for our study, with the $B - V$ colour index, the used S$_{\rm{MWO}}$-value, and the
rotational period. Also, for stars taken from the catalogue by \cite{wright2011ApJ.743.48W},
the adopted value of L$_{\rm{X}}$/L$_{\rm{bol}}$ value is provided.

\section{R$^{+}_{\rm{HK}}$: The magnetic activity related part of the Ca II H\&K emission}
\label{sec_rhkplus}

\subsection{Calculation of the R$^{+}_{\rm{HK}}$ index}

To investigate the rotation-activity connection we
used the chromospheric flux excess, i.e. the purely
activity related part of the chromospheric Ca II H\&K flux 
F$_{\rm HK}$, in the past often referred to as excess flux.  
For this purpose, we first had to convert the measured
S$_{\rm{MWO}}$ index into
physical surface fluxes F$_{\rm HK}$, which still contain three
contributions, i.e. emission from the photosphere, a so-called basal flux present
even in the total absence of magnetic activity, and finally, the magnetic
activity related emission in the line cores of the Ca II~H\&K absorption lines.

\cite{middelkoop1982} and \cite{rutten1984} were the first to propose 
a method to convert the S$_{\rm{MWO}}$ index into an absolute flux.
To estimate the true chromospheric flux excess, i.e. the flux without any
photospheric contribution, the photospheric flux component must be removed.  
To do so, \cite{linsky1979}, and later \cite{noyes1984} introduced 
the R$^{'}_{\rm{HK}}$ index  by defining
\begin{eqnarray}
\label{eq_rhk}
R^{'}_{\rm{HK}} & = & \frac{F_{\rm HK}-F_{\rm phot}}{F_{\rm{bol}}},
\end{eqnarray}
where $F_{HK}$ denotes the flux in the Ca~II~H\&K lines, 
$F_{phot}$ the photospheric flux contribution in the spectral range of
the Ca~II~H\&K lines, and $F_{\rm{bol}}$ the bolometric stellar flux. 

In a second step, we must subtract those contributions to the
chromospheric Ca II H\&K flux $F_{\rm HK}$ that are 
not related to magnetic activity.  This issue was addressed by
\cite{mittag2013A&A549A117M}, who defined a new chromospheric 
flux excess index, or purely magnetic activity index R$^{+}_{\rm{HK}}$, through
\begin{eqnarray}
R^{+}_{\rm{HK}} & = & \frac{F_{\rm HK}-F_{\rm phot}-F_{\rm basal}}{F_{\rm{bol}}},
\end{eqnarray}
where the quantities $F_{\rm HK}$,  $F_{\rm phot}$, and $F_{\rm{bol}}$ have the same 
meaning as in Eq.~\ref{eq_rhk}.
In addition, $F_{basal}$ denotes the so-called 
basal chromospheric flux in the Ca~II~H\&K lines \citep{Schrijver1987}, which is present in all 
stars, even in the most inactive stars, and 
cannot be related to any form of dynamo-driven magnetic activity.

In this R$^{+}_{\rm{HK}}$ index, both the photospheric flux component and basal chromospheric flux are subtracted as precisely as possible using PHOENIX
model atmospheres \citep{Hauschildt1999} to quantify the photospheric line core fluxes and
modern measurements of the basal flux. Consequently, the R$^{+}_{\rm{HK}}$ index
ought to represent the genuine, magnetic activity related 
chromospheric flux excess, i.e. the pure
contribution of magnetic activity to the chromospheric Ca~II~H\&K
emission.  For this reason, we only used the R$^{+}_{\rm{HK}}$ index and calculated its values following \cite{mittag2013A&A549A117M}, who give a detailed 
description of the chosen approach.

\subsection{How well does the R$^{+}_{\rm{HK}}$ index represent magnetic activity?}
\label{comp_r_hk_xray}

To demonstrate the usefulness of our pure representation of magnetic activity 
in the Ca~II~H\&K emission using the R$^{+}_{\rm{HK}}$ index, we show
how well this index relates the Ca II H\&K fluxes to the
X-ray fluxes in a sample of late-type stars; X-ray emission, in turn, is thought to
be the most direct proxy for magnetic activity that is not contaminated by
any photospheric contributions.

\begin{figure}  
\centering
\includegraphics[angle=0,scale=0.27]{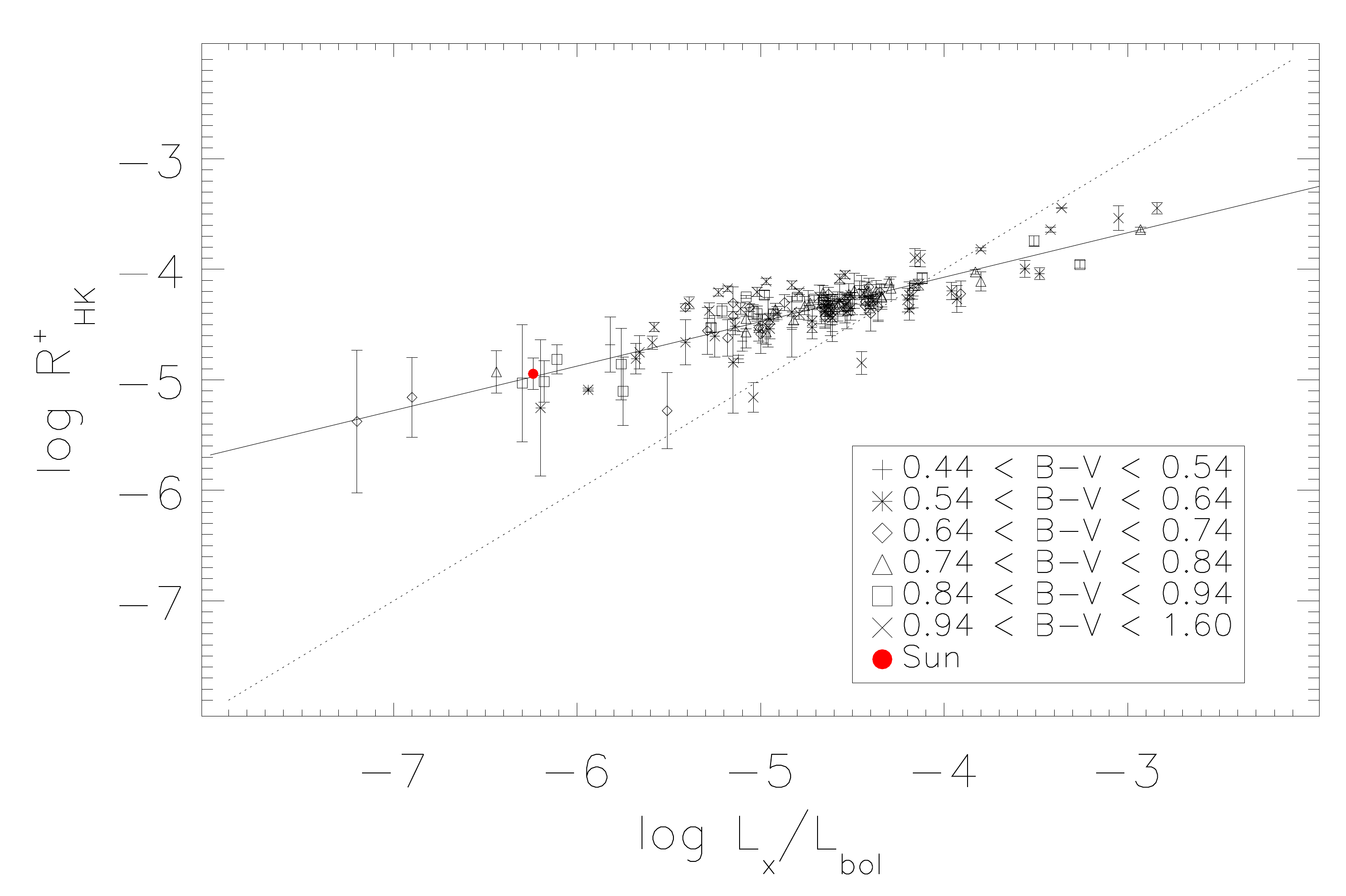}
\caption{Log R$^{+}_{\rm{HK}}$ vs. log L$_{\rm{X}}$/L$_{\rm{bol}}$ for the selected sample
star. The solid line indciates the best fit linear regression between
log R$^{+}_{\rm{HK}}$ and log L$_{\rm{X}}$/L$_{\rm{bol}}$ and the dotted line denotes
the identity. The solar values are indicated with a red dot; see text for details.
}
\label{compare_lx_r_hk}
\end{figure}

\cite{Maggio1987}, \cite{Schrijver1992}, and \cite{Hempelmann2006A&A} among others 
studied the correlation between the Ca~II~H\&K flux excess and X-ray activity. 
Using the catalogue compiled by \cite{wright2011ApJ.743.48W}, we selected 
156 stars with measured  rotational periods and X-ray fluxes for which
R$^{+}_{\rm{HK}}$ data is also available, albeit the X-ray and Ca~II 
measurements were not performed simultaneously.
In Fig.~\ref{compare_lx_r_hk} we plot log R$^{+}_{\rm{HK}}$ index data versus 
log L$_{\rm{X}}$/L$_{\rm{bol}}$ 
(all taken from \cite{wright2011ApJ.743.48W}) and recognize 
a clear correlation between these quantities. To estimate
the significance of this correlation we use Spearman's $\rho$ and obtain a
correlation coefficient of 0.76 with a significance
of 5.64$\cdot10^{-31}$. Finally, we determine the linear regression via 
a least-squares fit between both indicators and find
\begin{eqnarray}\label{eq_lx_rhk}
\log R^{+}_{\rm{HK}} = (-2.46\pm0.09) + (0.40\pm0.02) \log L_{\rm{X}}/L_{\rm{bol}}
\end{eqnarray}
with a standard deviation of 0.4; this relation is also plotted in 
Fig.~\ref{compare_lx_r_hk} as a solid line.  We note that the
slope of the relation is different from unity; the $R^{+}_{\rm{HK}}$ index values
span about two orders of magnitude; the $L_{\rm{X}}$/L$_{\rm{bol}}$ ratios span about
four orders of magnitude.  We also note that the Sun, indicated with a red dot 
in Fig.~\ref{compare_lx_r_hk}, fits perfectly to the observed distribution. We note
that the solar S-index values were taken from  \cite{b95} and the X-ray related values were taken from
\cite{wright2011ApJ.743.48W}.
 
\section{Empirical and theoretical convective turnover times}
\label{sec_convover}

\subsection{Physical meaning and empirical origins
     of the convective turnover time}

Naively one expects to find a clear correlation between the mean
activity and rotation period.  Yet, in 
their study of chromospheric emission \cite{noyes1984} pointed out that 
there is no such one-to-one relationship between the (logarithm of the)
R$^{'}_{\rm{HK}}$ index and the (logarithm of the) rotation period $P$.
However, by considering the colour and thus mass dependent Rossby number $R_{\rm{O}}$, 
defined as the ratio of the rotation period ($P$) to convective turnover 
time ($\tau_{c}$), i.e. $R_{\rm{O}} = P/\tau_{c}$, \cite{noyes1984} were able to demonstrate that the 
scatter in the rotation-activity relations can be significantly 
reduced and that the desired tight correlation between activity and 
rotation can indeed be produced. 

\cite{noyes1984} empirically introduced the convective turnover time $\tau_{c,days}$,
using the following representation by a function of stellar colour:
\begin{eqnarray}
\label{noyes_c_o_t}
\log \tau_{c,days} & = & 1.362-0.166x+0.025~x^{2}-5.323~x^{3}
\end{eqnarray}
for $x > 0$, and
\begin{eqnarray}
\log \tau_{c,days} & = & 1.362 -0.14~x
\end{eqnarray}
for $x < 0$, with x being defined as
\begin{eqnarray}
\label{def_x}
x = 1-(B-V),
\end{eqnarray}
and found that the relation
\begin{eqnarray}
\label{noyes_ross_number}
\log R_{O} & ~=~0.324-0.400y-0.283y^{2}-1.325y^{3},
\end{eqnarray}
with $y$ being defined as
\begin{eqnarray}
y = \log (R^{'}_{\rm{HK}}10^{5}),
\end{eqnarray}
results in minimal scatter in a R$^{'}_{\rm{HK}}$ versus $R_{O}$ diagram.  

In this fashion, the convective turnover time appears to provide the missing link between 
rotation period and activity, and its concept absorbs the colour and mass
dependence to provide unique activity-rotation relations. 
\cite{noyes1984} also used mixing length theory calculations in an attempt
to determine the convective turnover time versus spectral type from first 
principles.  

Inspecting Eq.~\ref{noyes_ross_number}, we note that
the parameter $R^{'}_{\rm{HK}}$ used by \cite{noyes1984} still includes the basal 
flux, hence $R^{'}_{\rm{HK}}$  does not vanish for a star with low activity 
near the basal flux level. 
This raises the question of what the activity-rotation relations 
look like at such low activity levels.
An inspection of Tab.~1 in \cite{noyes1984} shows that
the smallest reported value of $R^{'}_{\rm{HK}}$ (i.e. for the
star HD~13421) is $\approx$~6~$\times$10$^{-6}$ or $y \approx$ -0.205.
Evaluating Eq.~\ref{noyes_ross_number} with this value, we find a
ratio of rotation period over convective turnover time (defined 
according to the definition by \cite{noyes1984}) of $\approx$ 2.5. 
Such stars at the basal flux level, thus virtually without
activity, should then all have periods about $\approx$2.5 longer 
than their convective turnover time (again using the definition by 
\cite{noyes1984}).

\cite{rutten1987} investigated
the relationship between the Ca~II flux excess (which he called
$\log \Delta F^{~'}_{\rm{HK}}$)
and the rotation period in small $B - V$ intervals and found a good correlation
between these quantities. This fact was used by
\cite{stepien1989A&A...210..273S} to empirically define a convective turnover time.
However, he then rescaled his initial
relation by a factor of $\approx$2.7 to match the convective
turnover time defined by \cite{noyes1984}. 
Another empirically defined convective turnover time was introduced 
by \cite{2003A&A...397..147P}, who used X-ray data, yet followed
the same procedure as \cite{stepien1989A&A...210..273S}. The thus determined convective
turnover time was also rescaled to the convective turnover time for the Sun, which
was assumed to be 12.6 days.
Nevertheless, from \citet[][Figs. 10 and 11]{2003A&A...397..147P}, we
estimate a logarithmic shift of $\approx$0.8, which corresponds to a
scaling factor of $\approx$6.3 for the scaled and unscaled determined
turnover times.

A theoretical global convective turnover time was introduced 
by \cite{k-d1996ApJ...457..340K}, who argued that this should be considered
as the true convective turnover time; the same authors
denote the definition given by \cite{noyes1984} as the local convective turnover time because for this relation only one-half of the 
mixing length is used. Obviously, local and global convective turnover 
time differ only by a factor \citep{k-d1996ApJ...457..340K}.

\begin{figure*}[!t]
\centering{\includegraphics[angle=90,scale=0.6]{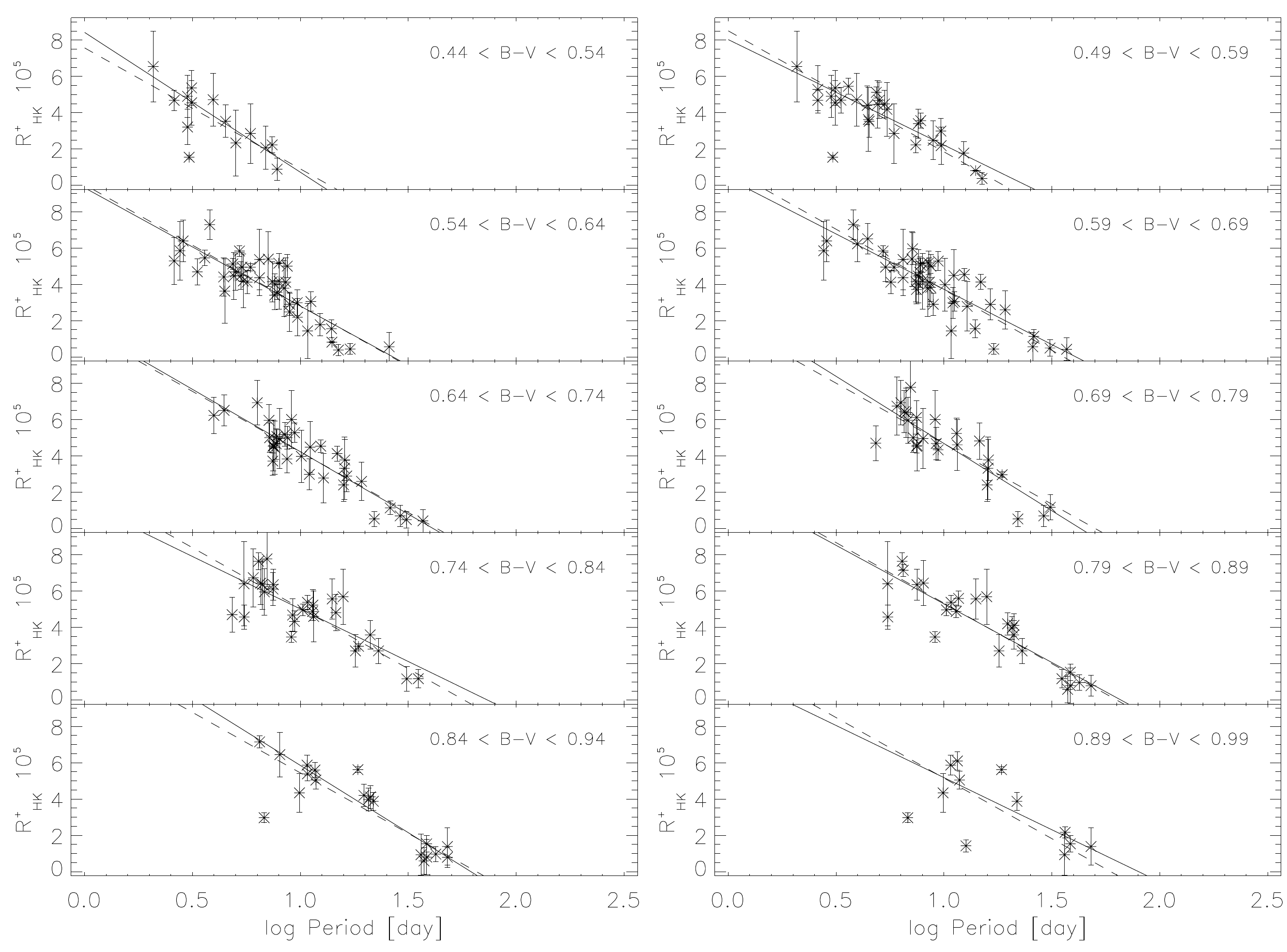}}
\caption{Pure activity flux excess index R$^{~+}_{\rm{HK}}$ values 
are plotted over the logarithm of the rotation periods in
adjacent $B - V$ intervals. The solid lines show the linear fit 
to each such data set.
The dashed lines depict the linear fit with the mean
slope that is used to estimate the convective turnover time.}
\label{flux_period_trents}
\end{figure*}

This can be confirmed by comparing the theoretical values for 
the local and global convective 
turnover times computed by \cite{landin2010A&A...510A..46L}.
According to their calculations we find that this factor between 
local and global 
convective turnover times is, on average, $\approx$ 2.4 for  
ages larger than 100  Myears, where a constant age dependency 
may be assumed. The factor of 2.4 is very much comparable 
with the factor of 2.5 that we find from our above discussion of
rotation periods of inactive stars
and also with the factor
derived above by \cite{stepien1989A&A...210..273S}. Consequently, 
we may indeed assume that the aforementioned envelope of the period versus $B - V$
distribution is given by the global convective turnover time (thereafter, 
referred to as convective turnover time $\tau_{\rm{c}}$).

\subsection{Rotation periods of inactive stars and the global convective turnover time}

\subsubsection{Basal flux timescale}

From the above discussion it clearly appears advantageous to use an activity measure
that explicitly excludes the basal flux as does the
$R^{~+}_{\rm{HK}}$ index. Because of its definition as a purely activity related
quantity, it should vanish for completely inactive stars. 
We are therefore led to describe the activity-rotation relationship in the form
\begin{eqnarray}
\label{general_a-r-relation}
\log P [day] & = & \log \tau_{c}(B - V)~[day] + f(R^{+}_{\rm{HK}}),
\end{eqnarray}
where the quantity $\tau_{c}$ is a colour-dependent timescale interpreted 
--~for the time being~--
as the rotation period of a star without any magnetic activity and 
$f(R^{+}_{\rm{HK}})$ is 
--~again for the time being~-- an arbitrary function, which is
zero when $R^{+}_{\rm{HK}}$ becomes zero. 

We next consider only less active stars with small flux excesses
(i.e. R$^{+}_{\rm{HK}}10^{5}\le$ 8) and
therefore linearly approximate the function $f(R^{+}_{\rm{HK}})$ introduced
in Eq.~\ref{general_a-r-relation}
through
\begin{eqnarray}
\label{functional_form}
f(R^{+}_{\rm{HK}}) = -~A \times R^{+}_{\rm{HK}}10^{5}
\end{eqnarray} 
with some constant $A$, universally
valid for all types of cool main-sequence stars.  

Following \cite{stepien1989A&A...210..273S}, we then subdivide 
our sample stars into smaller colour ranges listed in
Tab.~\ref{tabjs}. The same table also provides the number of stars in
each colour bin; we note that we use overlapping bins to ensure the
availability  of a sufficient number of stars in each colour bin with
the consequence that our fit results are not independent. For each of the
bins listed in Tab.~\ref{tabjs} we then perform a linear regression 
via a least-squares  fit between the R$^{+}_{\rm{HK}}10^{5}$-values and the
logarithmic rotation periods, where 
the slopes of these regressions are the inverse values of the 
slope A in Eq.~\ref{functional_form}.  In Fig.~\ref{flux_period_trents}
we plot the data and the so determined linear regressions for those
$B - V$ ranges with more than ten objects in the used colour range. 
Finally, the values of A are averaged 
and we obtain a general slope A of -0.15 with a standard 
deviation of 0.02. With this slope A the logarithmic basal flux timescales $\tau_{c,B-V bin}(B-V)$ are estimated for each $B - V$ colour bin 
listed in Tab.~\ref{tabjs}.

\begin{table}
\caption{Results of the convective turnover time estimation for each $B~-~V$ bin}
\begin{center}
\begin{tabular}{c c c}
  \hline \hline
  \noalign{\smallskip}
$B-V$ & Number   & $\log \tau_{c,B-V bin}$ \\
range & of used stars & \\
\hline
\noalign{\smallskip}
0.44-0.54 & 13 & 1.17$\pm$0.08\\
0.49-0.59 & 27 & 1.30$\pm$0.05\\
0.54-0.64 & 40 & 1.42$\pm$0.05\\
0.59-0.69 & 44 & 1.55$\pm$0.05\\
0.64-0.74 & 33 & 1.63$\pm$0.05\\
0.69-0.79 & 25 & 1.70$\pm$0.08\\
0.74-0.84 & 27 & 1.76$\pm$0.10\\
0.79-0.89 & 24 & 1.80$\pm$0.07\\
0.84-0.94 & 19 & 1.84$\pm$0.05\\
0.89-0.99 & 11 & 1.82$\pm$0.14\\
0.94-1.14 & 7  & 1.95$\pm$0.30\\
1.04-1.24 & 4  & 2.05$\pm$0.35\\
1.14-1.44 & 6  & 2.04$\pm$0.23\\
1.24-1.54 & 10 & 2.14$\pm$0.19\\
1.40-1.60 & 8  & 2.14$\pm$0.21\\
\hline
\end{tabular}
\label{tabjs}
\end{center}
\end{table}

\begin{figure}
\includegraphics[angle=90,scale=0.3]{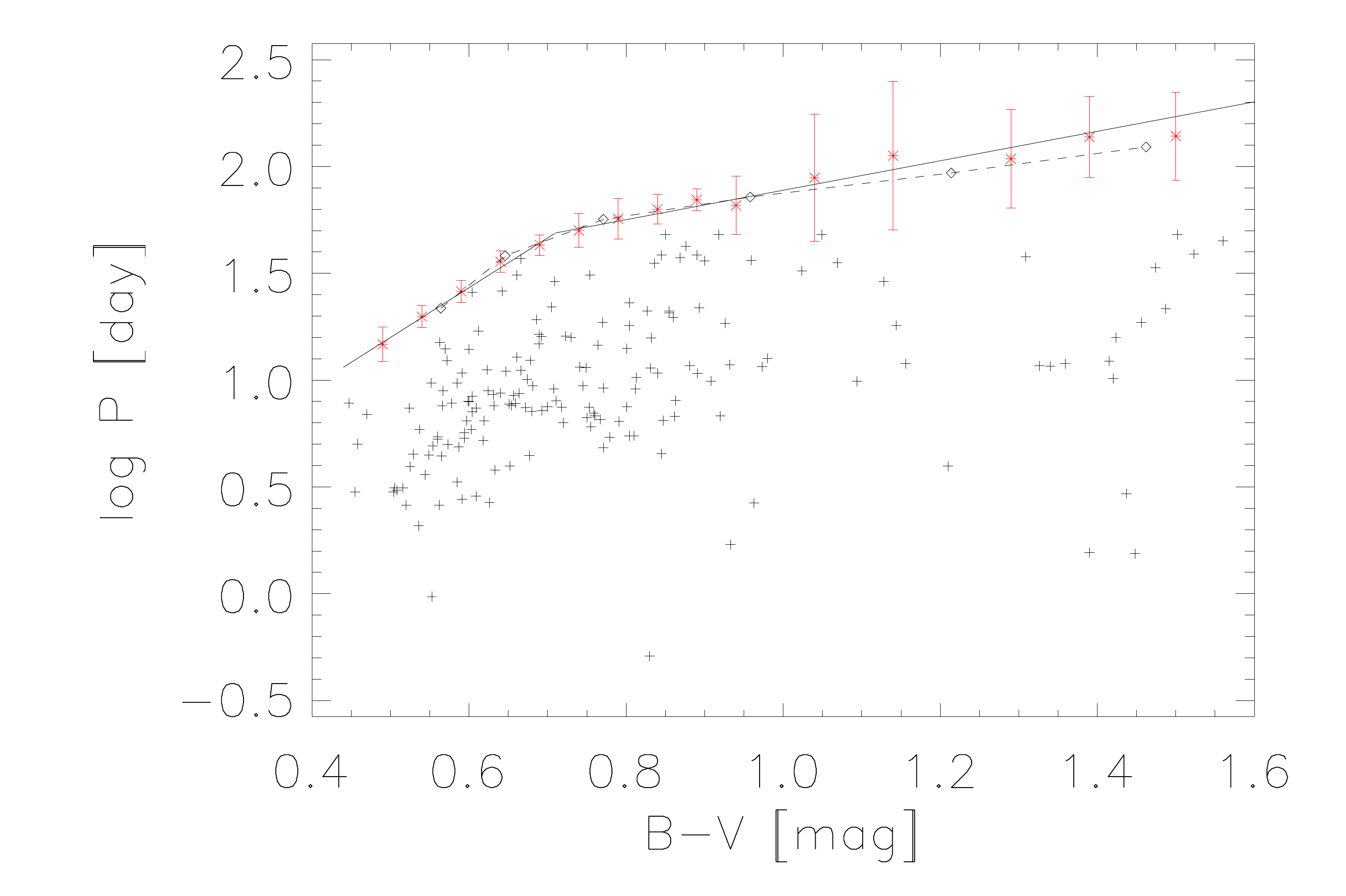}
\caption{Logarithm of the rotation periods vs. $B - V$. 
The red data points shown our $\tau_{c,B-V bin}$ with error.
The diamonds linked with a dashed line show the global 
convective turnover times derived by \cite{landin2010A&A...510A..46L}. The black solid
line shows the $\tau_{c}$ from Eq. \ref{eq_global_tau_a} and  \ref{eq_global_tau_b}}
\label{limit_periods}
\end{figure}

In Fig.~\ref{limit_periods} we plot for our whole sample the
measured (logarithmic) rotation periods versus $B-V$-colour and
additionally the derived basal timescales $\tau_{c,B-V bin}$ 
as listed in Tab.~\ref{tabjs}. We also plot the global
convective turnover times for stars with an 
age of 4.55~Gyrs derived by
\cite[][see Table 2]{landin2010A&A...510A..46L} as diamonds and to
guide the eye the points are connected by a dashed line; we note in this
context that the global convective turnover times derived by
\cite[][see Table 2]{landin2010A&A...510A..46L}
change only a little between ages 0.2 to 15 Gyr.

Fig.~\ref{limit_periods} demonstrates that, first, the colour dependent basal flux
timescales $\tau_{c,B-V bin}$  provide an upper envelope of the observed period
distribution of our sample stars, and second that the global
convective turnover times calculated by \cite{landin2010A&A...510A..46L}
appear to agree reasonably well with this upper period envelope.
We are therefore led to identify, based on this empirical result, 
the basal flux timescale $\tau_{c,B-V bin}$ with the rotation periods of
least active stars, and furthermore, that timescale with the global
convective turnover time. 

We now use this result to describe the latter empirically.
For practical reasons we use a simple linear fit to our values of
$\tau_{c,B-V bin}$ and obtain our final expression 
for the dependence of the convective turnover time $\tau_{c}$ versus $B-V$-colour in the form\\
\begin{eqnarray}
\label{eq_global_tau_a}
\log \tau_{c} = (1.06\pm0.07) + (2.33\pm0.37) ((B-V) - 0.44)
\end{eqnarray}
for stars with colours in the range $0.44 \le B - V \le 0.71$, and
for stars with $B - V > 0.71$ 
\begin{eqnarray}\label{eq_global_tau_b}
\log \tau_{c} = (1.69\pm0.12) + (0.69\pm0.13) ((B-V) - 0.71),
\end{eqnarray}
where the errors provide an estimate of the uncertainties involved.
These relations are plotted as a solid line in Fig.\ref{limit_periods}. 

\begin{figure}
\centering
\includegraphics[scale=0.3]{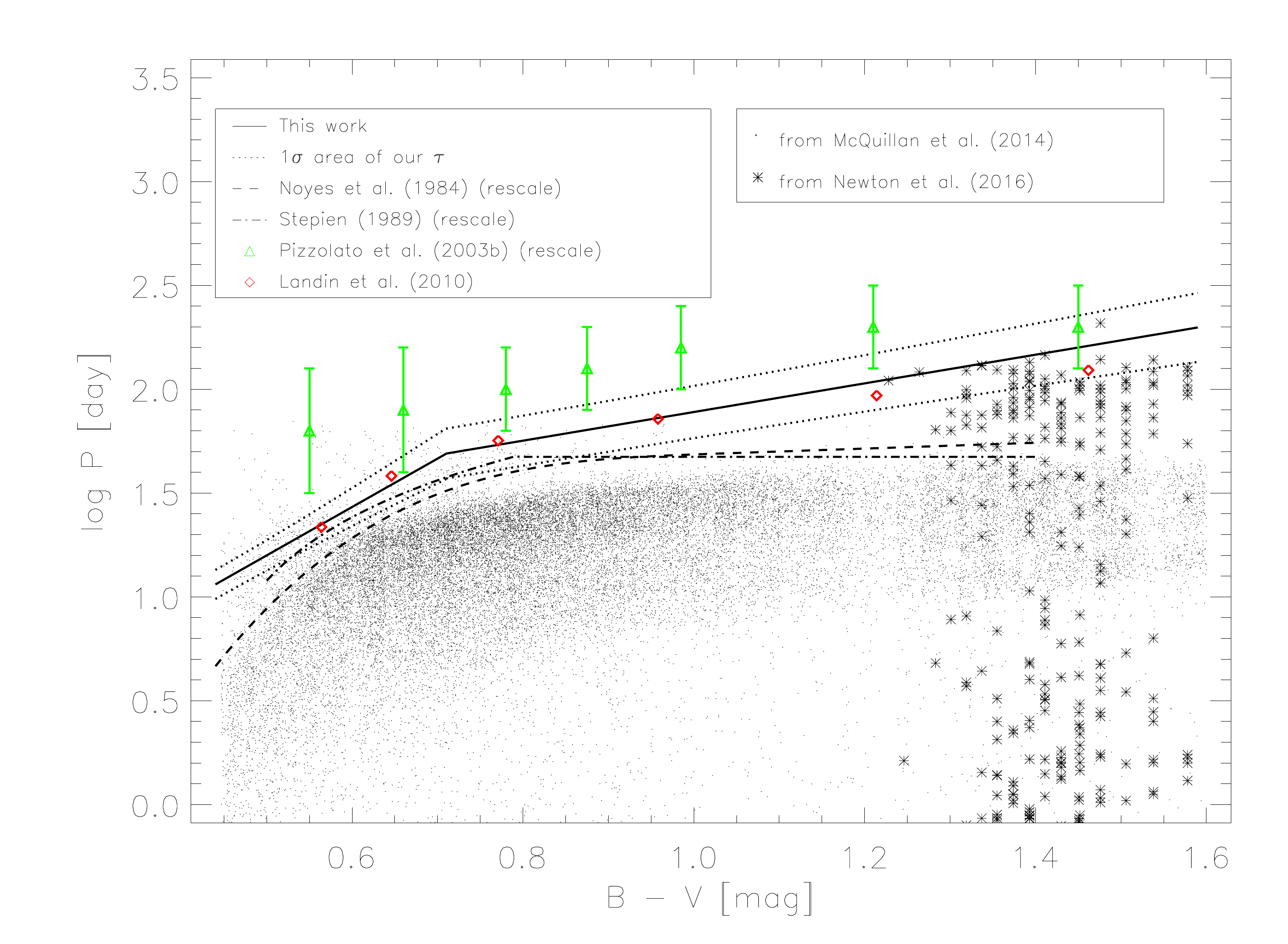}
\caption{Logarithm of the rotation periods vs. $B - V$ from the catalogue
  by \cite{mcquillan2014yCat} and \cite{newton2016yCat}.
Furthermore, our empirical convective turnover time calculated with
Eq. \ref{eq_global_tau_a} and \ref{eq_global_tau_b} is plotted as a
solid line. The dotted line is labelled as the 1$\sigma$ error area 
of convective turnover time. The rescaled relations by
\cite{noyes1984} and \cite{stepien1989A&A...210..273S} are plotted 
as dashed and dashed dotted line, respectively. The green 
triangle points are labelled as the values of the empirical convective turnover time
by  \cite{2003A&A...397..147P} and the 
values by \cite{landin2010A&A...510A..46L} are labelled with red diamonds.}
\label{kepler_periods}
\end{figure}

\subsubsection{Convective turnover time as upper envelope of P$_{rot}$ versus $B - V$ distribution?} 
\label{upper_envel_test}

Recently large samples of stars with measured rotation periods have become available
as a result of space-based and ground-based surveys.  Specifically, 
\cite{2014ApJS..211...24M} presented period measurements from
high-precision {\it Kepler} data for 34030 late-type stars. 
In their catalogue \cite{mcquillan2014yCat} give T$_{\rm{eff}}$ values 
rather than $B -V$ values; to convert into B-V colours we use
the T$_{\rm{eff}}(B - V)$ relation from \citet[Eq. 14.17]{gray2005oasp.book.G} and
finally select 29534 stars with $B - V$ colours between 0.4 and 1.6 and 
a log~g value greater than 4.2. In this context, not all of the period measurements
by \cite{mcquillan2014yCat} may necessarily refer to the stellar rotation period, however,
the bulk of the period data should really provide a good measure of rotation.

Next, \cite{Newton2016ApJ} presented rotation periods of 387 M-type stars derived from 
photometric MEarth observations. Since their catalogue \citep{newton2016yCat} does not provide $B-V$ 
colours, we obtained
the $B-V$ colours by cross-matching this catalogue with the LSPM-North catalogue by 
\cite{Lepine2005yCat} to obtain the J-K colours, which are then transformed
into $B - V$ colours by interpolation
using the colour table for solar metallicity
and log~g=4.5 by \cite{Worthey2011}. 
Finally, we selected 314 stars satisfying $B - V \le 1.6$. 

In total, we selected 29848 stars with measured rotation periods 
from both catalogues and plot these periods versus the derived
$B - V$ colours in Fig.~\ref{kepler_periods}; we also plot
our independently derived convective turnover time as a 
solid line; for further comparison
we also plot the rescaled relations by \cite{noyes1984}, \cite{stepien1989A&A...210..273S}, 
and \cite{2003A&A...397..147P} in Fig.~\ref{kepler_periods} with a dashed line, 
dash-dotted line, and green triangles, respectively.  
The theoretical global convective turnover time for 4.55 [Gyr] by
\cite[see table 2]{landin2010A&A...510A..46L} 
is depicted as red diamonds in Fig.~\ref{kepler_periods}; since
\cite{landin2010A&A...510A..46L} used $\log T_{\rm{eff}}$, we again compute
the respective $B - V$ values using the T$_{\rm{eff}}(B - V)$ relation 
from \citet[Eq. 14.17]{gray2005oasp.book.G}. 

Fig.~\ref{kepler_periods} demonstrates that our newly derived 
convective turnover time also provides
a good description of the upper envelope of the distribution of rotation periods 
of the stellar periods derived by \cite{mcquillan2014yCat} and \cite{newton2016yCat}.
We therefore argue that the convective turnover time can indeed be assumed 
to represent  an upper envelope of a period versus $B - V$ distribution.  From 
Fig.~\ref{kepler_periods} it is easily recognized that the rescaled  empirical, theoretical global, 
and our convective turnover time are located at the upper edge of the observed
period versus $B - V$ distribution with only small differences and therefore
conclude that our assumption that the true convective
turnover time is an upper envelope of the  
period versus $B - V$ distribution is indeed reasonable.

\subsubsection{Comparison to previously derived convective turnover times}
\label{sec_comp}

In the following we compare our new convective turnover time with previously 
derived convective turnover times.  
Comparing the different shapes of the convective turnover times, 
Fig.~\ref{kepler_periods} demonstrates that all relations show 
a strongly decreasing convective turnover time with smaller $B - V$ values and a more or less linear
behaviour at larger $B - V$ values, yet the
convective turnover times differ in their slopes for larger $B - V$ values. 
The slope of the theoretical convective turnover time is slightly smaller than that of
our convective turnover time, yet they still agree reasonably well.  In contrast, 
the slopes of the relations derived by \cite{noyes1984}, \cite{stepien1989A&A...210..273S}, 
and \cite{2003A&A...397..147P} are clearly lower, and  \cite{stepien1989A&A...210..273S} even
finds a constant convective turnover time with increasing $B - V$, similar to the findings of
\cite{2003A&A...397..147P}.   Furthermore, we note in context that the (unscaled) convective turnover
time derived by \cite{2003A&A...397..147P} (see Fig.~10 in this paper) exceeds the corresponding
estimates by \cite{k-d1996ApJ...457..340K} by about a factor of 2; on the other hand,
the estimates by \cite{k-d1996ApJ...457..340K} and \cite{landin2010A&A...510A..46L}
agree fairly well, so that the nature of this shift is entirely unclear.

The period distribution derived from the {\it Kepler} data by \cite{mcquillan2014yCat} 
shows a rather sharp edge at $\approx$ 40~days for stars with $B - V >$ 1.0.
This apparent edge lies well below our the convective turnover time, however, it is
consistent with the trends of the classical rescaled relations by \cite{noyes1984} 
and \cite{stepien1989A&A...210..273S}, which result in a nearly constant or constant 
convective turnover time for stars $B - V >$ 1.0.  
However, adding the M star data by \cite{newton2016yCat}, the gap between 
the data by \cite{mcquillan2014yCat} and our convective turnover time is filled up
and makes the relations of \cite{noyes1984} and \cite{stepien1989A&A...210..273S} 
inconsistent with the data. 

Finally, we compare the convective turnover time of the Sun calculated with our relation
with theoretical values listed in \citet[Fig. 16]{Spada2013ApJ}. These
theoretical values of the convective turnover time for the Sun range from
31.70 days to 42.12 days;
with our relation Eq. \ref{eq_global_tau_a}, we compute a convective turnover time
of 33.9$\pm$8.0 days, i.e. a value well in the range of the theoretical values.

\begin{figure}
\includegraphics[angle=0,scale=0.5]{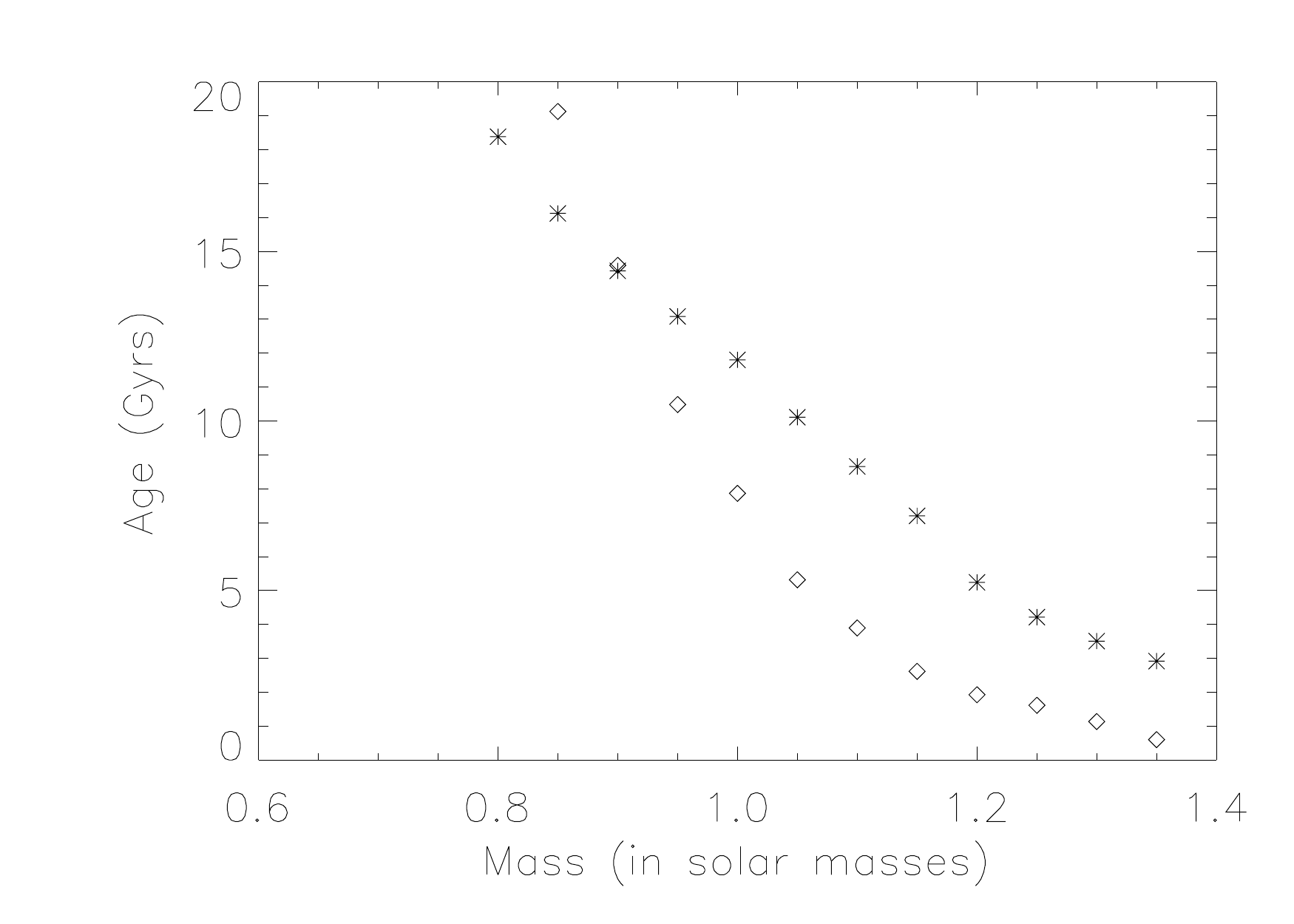}
\caption{Convective turnover time (asterisks) vs. main-sequence lifetime
(diamonds) vs. stellar mass.}
\label{fig_age}
\end{figure}

\subsection{Can stars rotate more slowly than their convective turnover times~?}

In the previous sections we introduced the convective turnover time empirically
as the upper envelope in the observed rotational period distribution of
late-type stars.  What remains somewhat open is why this upper envelope
exists in the first place.  After all, the observed rotational period 
distribution might be biased since it becomes increasingly difficult to measure rotation
periods for less and less active stars, and there is no physical law
preventing stars from braking to essentially zero rotational velocity.
We, however, argue that this problem does not appear in practice.
In gyrochronology the following equation is used to relate the stellar age
$t_{age}$ with the rotation period $P$ and the convective turnover time 
$\tau$ \citep{barnes2010}:
\begin{equation}
\label{eq_gyro}
t_{age} = \frac{\tau}{k_C} ln(\frac{P}{P_0}) + \frac{k_I}{2 \tau} (P^2-P_0^2),
\end{equation}
where the parameters $P_O$, $k_C$, and $k_I$ take on the values
$P_0$ = 1.1 day, $k_C$ = 0.646 day/Myr, and $k_I$ = 452 Myr/day.
Using Eq.~\ref{eq_gyro} we can compute that age $t_{break}$, when the 
rotation period $P$ equals the convective turnover time $\tau$, and
the resulting values are shown in Fig.~\ref{fig_age} as a function of
stellar mass.  

These values of $t_{break}$ need to be compared to the main-sequence
lifetimes $t_{ms}$. To estimate $t_{ms}$, we use the time span
between a star's arrival on the main sequence and its first turning point,
i.e. the time when the hydrogen fuel becomes exhausted in the core,
the star becomes redder, and starts ascending the giant branch.
We use the evolutionary tracks computed and tested by \cite{schroder1998} to
obtain the reliable main-sequence lifetimes 
$t_{ms}$ of the stars in the relevant mass range; the models
specifically deal with convective overshooting and its onset at about 
1.5$M_{\odot}$.

The so-computed values of $t_{ms}$ are also shown in Fig.~\ref{fig_age} 
(with diamonds). As becomes clear from Fig.~\ref{fig_age},  
for stars with masses $M_{*} > 0.8 M_{\odot}$ the
main-sequence lifetimes $t_{ms}$ are always shorter 
than the ages $t_{break}$.
In other words, stellar evolution, which makes the stars evolve and
move up the giant branch, is faster than magnetic braking, which
tries to spin down the stars to their convective turnover times.
For stars with masses $M_{*} < 0.8 M_{\odot}$ the main-sequence lifetimes $t_{ms}$ exceed the ages $t_{break}$, 
however, given the age of the universe, such stars have not had their chance
to experience rotational braking throughout their full main-sequence careers.
For such stars $t_{break}$ scales as a multiple
of $t_{ms}$, and as already noted by \cite{schroder2013} and
consistent with the studies by \cite{Reiners2012}, this
implies that magnetic braking scales with the relative main-sequence evolutionary age of solar-like stars of different masses. 
As a consequence, we infer that a complete slowdown of stellar
rotation by magnetic braking seems to be impossible on the main sequence
for stars like our Sun. 

Yet there is an additional argument in this context.  Once
the Rossby number reaches the value of unity and hence any chromospheric 
emission is reduced to the basal flux level, there is no magnetic dynamo
action with emerging active regions any more, rather only small-scale
activity is expected to be present on the star.  Following
models of the global magnetic field of the Sun under such conditions
\citep{meyer2016}, we speculate that magnetic braking becomes 
less efficient because of the decrease of the magnetic field strength,
which provides another reason why the rotation periods of basal flux stars 
form an envelope to the observed rotation periods.

\section{Magnetic activity related Ca~II~H\&K emission and Rossby number}

\subsection{R$^{+}_{\rm{HK}}$ versus Rossby number}
The Ca~II~H\&K excess flux is thought to be caused purely by dynamo-produced
stellar magnetic activity.  An important ingredient of dynamo action is of course
rotation, which is directly involved in the definition of the Rossby number. 
The correlation between Ca~II H\&K excess flux 
and Rossby number was established long ago \citep[e.g. ][]{noyes1984}, and
the general shape of the dependence of Ca~II H\&K excess flux and
Rossby number has been derived by previous authors
(see e.g. Noyes et al. 1984, Fig. 8; Mamajek \& Hillenbrand 2008, Fig. 7).
However, the Rossby number definitions previously used 
do not constrain the Rossby number to a well-defined range. 
 By contrast, our approach of equating global convective
turnover times with the rotation periods of entirely inactive main-sequence stars leads to a well-defined Rossby number range 
between 0.0 and 1.0; the
latter value denotes the maximal rotation period a star can have. 
We now address the question of which range of  Ca II H\&K excess 
flux this range of the Rossby numbers corresponds to.

\begin{figure}
\centering
\includegraphics[angle=90, scale=0.3]{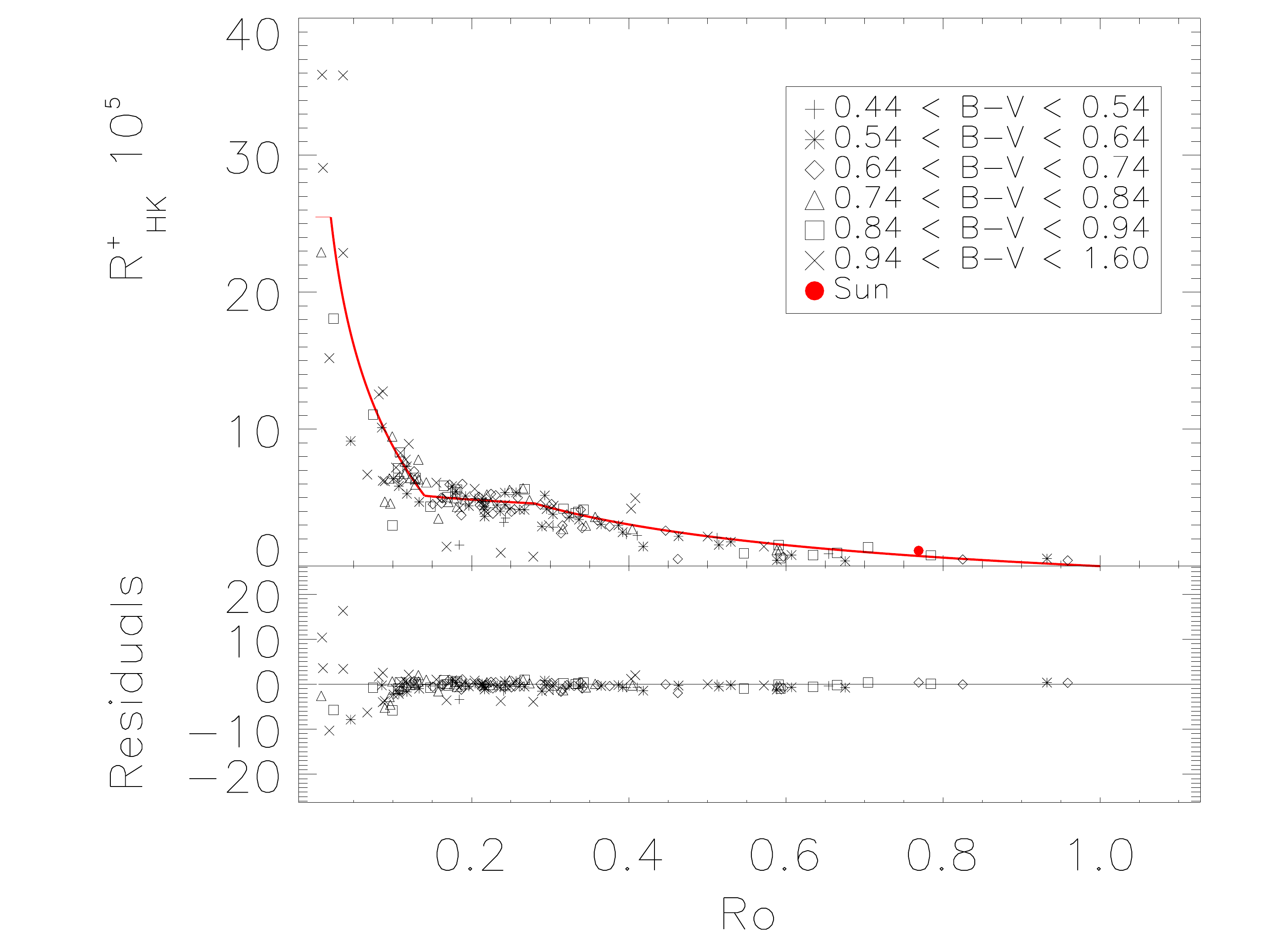}
\caption{Ca II H\&K flux excess R$^{+}_{\rm{HK}}10^{5}$ vs. Rossby
  number is shown in the upper plot. The lower plot
  shows the residuals of the flux trend.}
\label{flux_ro}
\end{figure}

In Fig. \ref{flux_ro} we plot the values of Ca II H\&K excess flux R$^{+}_{\rm{HK}}$ versus
Rossby number R$_{\rm{O}}$ computed from our empirically derived convective
turnover time (cf. Eq. \ref{eq_global_tau_a} and \ref{eq_global_tau_b}) for the objects 
listed in Tab.~\ref{data_tab}.  As expected, large values of R$^{+}_{\rm{HK}}$ go along
with small values of Rossby numbers, however, we also recognize a clear change in the
trend of this dependence. For fast rotators (with small R$_{\rm{O}}$) 
the excess flux decreases very fast from the saturation level to a flux level lower than 
R$^{+}_{\rm{HK}}10^{5} \approx$ 6, yet the
saturation level of R$^{+}_{\rm{HK}}10^{5}$ = 25.5 and log
R$^{+}_{\rm{HK}}$ = -3.59, respectively, 
cannot be particularly well estimated from our R$^{+}_{\rm{HK}}$ data because we have 
only seven data points in range lower than Ro=0.04. Furthermore, the Rossby number,
below which saturation sets in, is also not well determined from our data.

Following an initial strong decrease, the excess flux continues to decrease less steeply
until a value of R$_{\rm{O}} \approx0.28$ is reached. To estimate the trend in 
this range we ignore those four data points with a flux excess lower than 2
because these data points are located more than 3$\sigma$ below the rest and
test whether a linear trend is justified; from Spearsman's $\rho$ we 
estimate the correlation coefficient and its significance, which turn out
to be -0.23 with a significance of 92\%.   For larger values of R$_{\rm{O}}$, the flux
decrease becomes slightly stronger and the flux excess eventually
disappears with the Rossby number reaching unity; we also refer to 
Fig.~\ref{compare_h_alpha_lx_r_hk} (top panel) for a double logarithmic
plot of the same data.

To empirically describe the dependence of R$^{+}_{\rm{HK}}10^{5}$ on Rossby number we 
therefore introduce four parameter regions and we obtain the following approximations:\\
For the range Ro~$<$0.021 we find
\begin{eqnarray}\label{eq_ro_d}
\rm{R}^{+}_{\rm{HK}}10^{5} & \approx & 25.5
\end{eqnarray}
for the range 0.021$\le$Ro$<$0.14 we have
\begin{eqnarray}\label{eq_ro_c}
\rm{R}^{+}_{\rm{HK}}10^{5} & = &  (5.14\pm0.16) + (-24.55\pm0.14)x
\end{eqnarray}
with x = $\log$(Ro)-$\log$(0.14), and for 0.14$\le$Ro$<$0.28 we have
\begin{eqnarray}\label{eq_ro_b}
\rm{R}^{+}_{\rm{HK}}10^{5} & = &  (4.58\pm0.09) + (-1.87\pm0.42)x
\end{eqnarray}
with x = $\log$(Ro)-$\log$(0.28) and finally for Ro$\ge$0.28 we have
\begin{eqnarray}\label{eq_ro_a}
\rm{R}^{+}_{\rm{HK}}10^{5} & = &  (-6.03\pm0.17)x + (4.09\pm0.32)x^{2} 
\end{eqnarray}
with x = $\log$(Ro).

The relation described by Eq.~\ref{eq_ro_d}-Eq.~\ref{eq_ro_a} is shown as a
red solid line in Fig.~\ref{flux_ro}.  
We estimate a standard deviation of $\approx$ 2.3. However, the size of the 
scatter is different in the different ranges.  We obtain a standard deviation of the 
residuals of 8.8 for Ro$<$0.021, 3.8 for 0.024$\le$Ro$<$0.14,
1.1 for 0.14$\le$Ro$<$0.28, and 0.7 for Ro$\ge$0.28.
To test for a remaining colour dependency in the flux evolution, the data in different
$B - V$ ranges are labelled with different symbols in Fig. \ref{flux_ro} and no
colour dependence is seen.

\subsection{Comparison of R$^{+}_{\rm{HK}}$, H$\alpha$, and X-ray flux evolution versus Rossby number}
\label{sec_multi}

Finally, we compared our newly derived R$^{+}_{\rm{HK}}$ versus Rossby number relation with
the equivalent relations derived for H$\alpha$ and X-ray data.  For this comparison
we used the $L_{H{\alpha}}$/L$_{\rm{bol}}$-values derived by \cite{Newton2017yCat} for a sample
of M dwarfs and the L$_{\rm{X}}$/L$_{\rm{bol}}$ values derived by
\cite{wright2011ApJ.743.48W} for a sample of late-type stars; in all cases we only used stars with
measured rotation periods.

\begin{figure} 
\centering
\includegraphics[angle=90, scale=0.35]{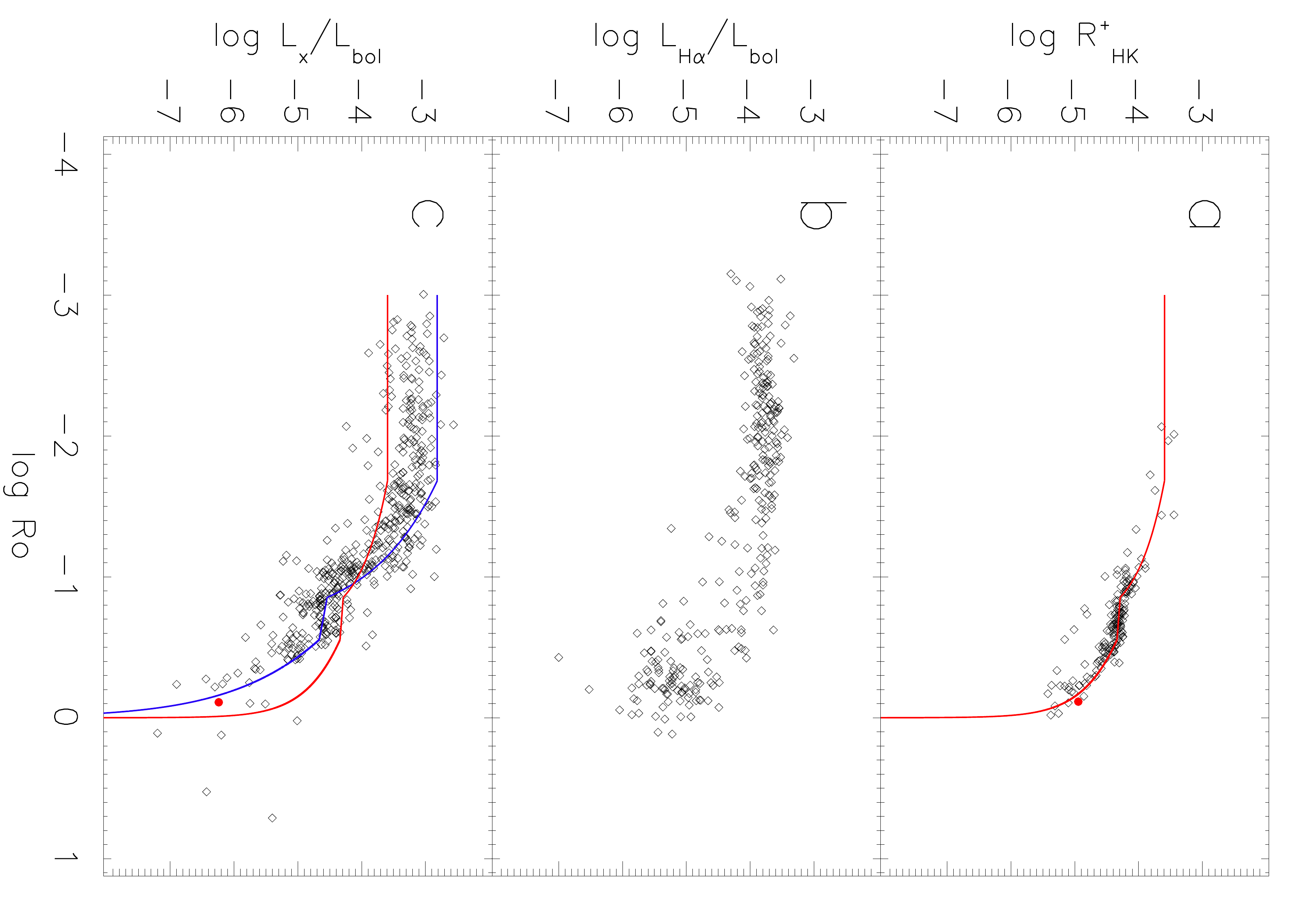}
\caption{Log R$^{+}_{\rm{HK}}$, log L$_{H\alpha}$/L$_{\rm{bol}}$, and log L$_{\rm{X}}$/L$_{\rm{bol}}$ vs. Rossby number are shown.
  The red line depicted our flux trend for R$^{+}_{\rm{HK}}$ vs. Rossby number and the red dot indicates the Sun. The blue line
indicates the R$^{+}_{\rm{HK}}$ trend transformed into the log L$_{\rm{X}}$/L$_{\rm{bol}}$ scale with Eq. \ref{eq_lx_rhk}.}
\label{compare_h_alpha_lx_r_hk}
\end{figure}

We calculated the Rossby numbers for the selected sample objects with our new
empirical convective turnover time. In both catalogues, no $B-V$-values 
are provided.
For the objects studied by \cite{Newton2017yCat}, we used the same
method as described in Sec.~\ref{upper_envel_test}.
The catalogue by \cite{wright2011ApJ.743.48W} provides
effective temperatures for all entries, which we transform into the
$B - V$ colour using the T$_{\rm{eff}}(B - V)$ relation from 
\citet[Eq. 14.17]{gray2005oasp.book.G}. 
In Fig.~\ref{compare_h_alpha_lx_r_hk} we plot the dependence of these three
activity indicators versus Rossby number.  Clearly, the general shape of the relation
is the same in all three indicators, yet some differences are discernible.
The saturation level in the log R$^{+}_{\rm{HK}}$ values  
and the log $L_{H{\alpha}}$/L$_{\rm{bol}}$ values
is comparable, considering the fact that we have only very few data 
points at the saturation level, while the
saturation level of the log L$_{\rm{X}}$/L$_{\rm{bol}}$ data
is higher than that of the log R$^{+}_{\rm{HK}}$ values.

To compare the Rossby number dependence of the various activity indicators 
we plot our flux trend Eq.~\ref{eq_ro_d} - \ref{eq_ro_a} into 
Fig.~\ref{compare_h_alpha_lx_r_hk} as a red line.  As is clear
from Fig.~\ref{compare_h_alpha_lx_r_hk}, the H$_{\alpha}$ data
show a large scatter for data with Rossby numbers larger 0.1, which
makes a comparison difficult. Therefore, we can only say that the
log L$_{H\alpha}$/L$_{\rm{bol}}$ dependence on Rossby number is described in general; there is a possible overprediction in the saturation region.
As far as the X-ray data are concerned, it is clear that the flux trend 
of Eq.~\ref{eq_ro_d} - \ref{eq_ro_a} does not fit the data particularly well.
However, using the transformation relation between log~R$^{+}_{\rm{HK}}$
and log~L$_{\rm{X}}$/L$_{\rm{bol}}$ derived in Sec.~\ref{comp_r_hk_xray}, 
we can rescale the activity-Rossby number relation and arrive at the
blue line in the Fig.~\ref{compare_h_alpha_lx_r_hk}c, which describes 
the logL$_{\rm{X}}$/L$_{\rm{bol}}$ versus Rossby number relation very well,
except in the saturation region, which lies clearly above the observed
X-ray saturation.
We therefore suspect that our saturation level could be too high, possibly 
as a consequence of the low number of data points to estimate the saturation level
in the Ca~II data.  

\section{Summary and conclusions}
\label{sec_conc}

In this paper we revisit the connection between the purely magnetic
activity related Ca~II~H\&K flux excess and the rotation period for main-sequence stars, focussing on the R$^{+}_{\rm{HK}}$ index. This index is determined by first subtracting all non-magnetic 
activity contributions from the Ca~II~H\&K line core flux, notably the
photospheric flux and basal chromospheric flux, and second 
normalizing by the bolometric flux. By contrast, the older activity 
index $R^{'}_{\rm{HK}}$ still includes the basal chromospheric flux and therefore
contains contributions not directly attributable to
the effects of magnetic activity.
We demonstrate that the R$^{+}_{\rm{HK}}$ index provides a very good physical 
representation of the stellar chromospheric activity, since it relates very 
well with the coronal X-ray flux, using the L$_{\rm{X}}$/L$_{\rm{bol}}$ data 
for a sample of stars with measured S-indices, X-ray fluxes, and rotation
periods. 

Revisiting the rotation-activity relation with our
new indicator we find that the chromospheric flux reaches zero for
inactive (basal flux) stars, which have a well-defined colour dependent
rotation period. This result is intuitively consistent
with the idea that in an inactive star no significant magnetic
braking can take place any further, which forces it to maintain 
its rotational angular momentum from that point in its evolution, until
substantial mass loss sets in.  These resulting basal flux rotation periods
only depend on the $B-V$ colour and compare to the data in the same way as 
earlier empirical concepts of convective turnover time.

We show that this convective turnover time provides a very good representation
of the upper envelope of the observed period versus $B - V$ distribution for a very
large sample of cool stars and then, using both by empirical and theoretical approaches, we
compare the resulting convective turnover times with previous 
prescriptions. 
Our approach turns out to be fully consistent with the theoretical 
relation proposed by \cite{landin2010A&A...510A..46L}, but we find
significant differences between our relation and those derived by
\cite{noyes1984}, \cite{stepien1989A&A...210..273S},
and \cite{2003A&A...397..147P}, especially in the slope
in the range  $B - V >$  $\approx$0.9. 

We emphasize that with this new definition of the convective 
turnover time, the Rossby number (for main-sequence stars) 
has a well-defined definition range from zero to unity.
When the relation of these new Rossby numbers with the Ca II H\&K flux
excess is investigated, a clear evolutionary picture emerges, 
dividing this relation into distinct parts: for small values of the 
Rossby number (fast rotating star with large activity) the excess 
flux is saturated. Then it initially decreases quickly and then more slowly, 
thus asymptotically reaching zero, where the Rossby number 
has reached unity. A good example is the well-known planet host
51 Peg (HD~217014), which is a somewhat
evolved main-sequence stars \citep{Mittag2016A&A...591A..89M} 
with a low S$_{\rm{MWO}}$ value 
of 0.149, very close to the basal S-index
level of 0.144 \citep{mittag2013A&A549A117M}.
With our new convective turnover time the Rossby number of 51 Peg is 0.96$\pm$0.24,
i.e. very close to unity and therefore its rotation period of 37~days 
\citep{wright2011ApJ.743.48W} should essentially comparable with the basal flux
timescale of 38.6~days computed with Eq.~\ref{eq_global_tau_a}.

Finally, we compare
the estimated trend of R$^{+}_{\rm{HK}}$ versus Rossby number with 
$L_{\rm{H}\alpha}$/L$_{\rm{bol}}$ and L$_{\rm{X}}$/L$_{\rm{bol}}$ data versus 
Rossby number and find that the general trend
is the same in all three activity indices. Furthermore, plotting 
our flux trend into the good match of the data is obtained especially with the X-ray data after the
adjustment of the trend at the X-ray data with the transformation 
equation, which suggests a universality of the Rossby number activity relations.

\begin{acknowledgements}
This research has made use of the VizieR catalogue access tool, CDS,
Strasbourg, France. The original description of the VizieR service was
published in A\&AS 143, 23.
\end{acknowledgements}


\newpage
\begin{appendix}
  \section{Table}
\begin{table*}
\caption{Objects used with the corresponding $B - V$ colour index, S-values, 
rotational period, and log L$_{\rm{X}}$/L$_{\rm{bol}}$ are listed. The log L$_{\rm{X}}$/L$_{\rm{bol}}$ values are from \citet{wright2011ApJ.743.48W}. }
\label{data_tab}
\smallskip
\begin{center}
\begin{small}
\begin{tabular}{l c c l c c c l}
\hline\hline
\noalign{\smallskip}
Object & $B-V$ & S & Ref. & Per [day] & Ref & log L$_{\rm{X}}$/L$_{\rm{bol}}$ \\
\hline
\noalign{\smallskip}
SUN          & 0.642 & 0.179$\pm$0.001 & B95                          & 26.09 & W11  & -6.24 \\
HD 691       & 0.755 & 0.473$\pm$0.057 & W04,W07,I10                  &  6.05 & W11  & -4.29 \\
HD 1326 A    & 1.560 & 0.567$\pm$0.023 & W04,W07,H14                  & 44.80 & H14  &       \\
HD 1835      & 0.659 & 0.349$\pm$0.001 & B95                          &  7.78 & W11  & -4.43 \\
HD 2454      & 0.447 & 0.170$\pm$0.001 & B95                          &  7.80 & R87  &       \\
HD 3651      & 0.850 & 0.176$\pm$0.001 & B95                          & 48.00 & W11  & -5.75 \\
HD 4628      & 0.890 & 0.230$\pm$0.001 & B95                          & 38.50 & W11  & -6.11 \\
HD 6963      & 0.730 & 0.246$\pm$0.022 & W04,W07,G03                  & 15.86 & W11  & -5.18 \\
HD 7590      & 0.594 & 0.293$\pm$0.008 & G03,W04,I10                  &  5.67 & W11  & -4.65 \\
HD 7661      & 0.753 & 0.440$\pm$0.025 & W04,G06,W07,S09,I10,A11      &  7.46 & W11  & -4.66 \\
HD 8907      & 0.505 & 0.293$\pm$0.028 & W04,W07,I10                  &  3.13 & W11  & -4.36 \\
HD 9472      & 0.666 & 0.326$\pm$0.035 & W04,W07,I10                  & 11.10 & W11  & -4.54 \\
HD 10476     & 0.836 & 0.198$\pm$0.001 & B95                          & 35.20 & W11  & -6.44 \\
HD 10780     & 0.804 & 0.280$\pm$0.001 & B95                          & 23.00 & W11  & -4.97 \\
HD 11507     & 1.424 & 2.039$\pm$0.136 & W04,G06                      & 15.80 & W11  & -4.83 \\
HD 11850     & 0.711 & 0.363$\pm$0.051 & W04,W07,I10                  &  8.00 & W11  & -5.15 \\
HD 12786     & 0.832 & 0.466$\pm$0.075 & W04,G06                      & 15.78 & W11  & -4.34 \\
HD 13507     & 0.672 & 0.294$\pm$0.021 & G03,W04,W07,I10              &  7.45 & W11  & -4.65 \\
HD 13531     & 0.700 & 0.341$\pm$0.029 & G03,W04,W07,I10              &  7.49 & W11  & -4.36 \\
HD 13579     & 0.920 & 0.340$\pm$0.019 & D91,W04,I10                  &  6.79 & W11  & -5.27 \\
HD 16157     & 1.390 & 6.689$\pm$1.600 & G06,S09                      &  1.56 & W11  & -3.05 \\
HD 16160     & 0.918 & 0.226$\pm$0.001 & B95                          & 48.00 & W11  & -5.76 \\
HD 16673     & 0.524 & 0.215$\pm$0.001 & B95                          &  7.40 & W11  & -5.10 \\
HD 17925     & 0.862 & 0.653$\pm$0.001 & B95                          &  6.76 & W11  & -4.12 \\
HD 18940     & 0.624 & 0.250$\pm$0.008 & W04,W07                      &  8.92 & W11  & -4.95 \\
HD 19019     & 0.552 & 0.215$\pm$0.013 & W04,W07,I10                  &  9.70 & W11  & -5.41 \\
HD 19632     & 0.678 & 0.332$\pm$0.013 & W04,G06,J06,W07              & 12.42 & W11  & -5.41 \\
HD 19668     & 0.810 & 0.492$\pm$0.028 & W04,W07,I10,J11              &  5.48 & W11  & -4.49 \\
HD 20630     & 0.681 & 0.366$\pm$0.001 & B95                          &  9.40 & W11  & -4.66 \\
HD 22049     & 0.881 & 0.496$\pm$0.001 & B95                          & 11.68 & W11  & -4.80 \\
HD 25457     & 0.516 & 0.323$\pm$0.019 & G03,W04,W07,S09,I10          &  3.13 & W11  & -4.17 \\
HD 25998     & 0.520 & 0.300$\pm$0.001 & B95                          &  2.60 & N84  &       \\ 
HD 26736     & 0.657 & 0.353$\pm$0.006 & D91,I10                      &  8.48 & W11  & -4.54 \\
HD 26767     & 0.640 & 0.340$\pm$0.026 & D91,W04                      &  8.69 & W11  & -4.40 \\
HD 26913     & 0.680 & 0.396$\pm$0.001 & B95                          &  7.15 & W11  & -4.63 \\
HD 26990     & 0.661 & 0.251$\pm$0.008 & W04,W07,S09,I10              & 12.80 & W11  & -5.29 \\
HD 27250     & 0.745 & 0.346$\pm$0.021 & D91,I10                      &  9.39 & W11  & -4.92 \\
HD 27406     & 0.560 & 0.289$\pm$0.001 & D91                          &  5.44 & W11  & -4.65 \\
HD 27836     & 0.604 & 0.345$\pm$0.011 & D91                          &  7.10 & W11  & -3.93 \\
HD 27859     & 0.599 & 0.296$\pm$0.012 & D91,W07,I10                  &  7.96 & W11  & -4.53 \\
HD 28068     & 0.651 & 0.329$\pm$0.032 & D91,W07                      &  7.73 & W11  & -4.43 \\
HD 28099     & 0.664 & 0.297$\pm$0.013 & D91,W07,I10                  &  8.67 & W11  & -4.62 \\
HD 28205     & 0.537 & 0.238$\pm$0.002 & D91                          &  5.87 & W11  & -4.83 \\
HD 28237     & 0.560 & 0.300$\pm$0.006 & D91,W04,I10                  &  5.30 & W11  & -4.55 \\
HD 28291     & 0.741 & 0.358$\pm$0.043 & D91,I10                      & 11.52 & W11  & -4.58 \\
HD 28344     & 0.609 & 0.297$\pm$0.020 & D91,W04,W07                  &  7.41 & W11  & -4.61 \\
HD 28495     & 0.759 & 0.529$\pm$0.051 & W04,I10                      &  7.00 & W11  & -3.80 \\
HD 28992     & 0.631 & 0.301$\pm$0.007 & D91,W07,I10                  &  8.55 & W11  & -4.83 \\
HD 29310     & 0.597 & 0.342$\pm$0.045 & D91,I10                      &  6.46 & W11  & -4.57 \\
HD 30495     & 0.632 & 0.297$\pm$0.001 & B95                          &  7.60 & W11  & -4.91 \\
HD 32147     & 1.049 & 0.286$\pm$0.001 & B95                          & 48.00 & W11  &       \\
HD 32850     & 0.804 & 0.280$\pm$0.020 & G03,W04,W07                  & 17.99 & W11  & -5.08 \\
HD 35296     & 0.544 & 0.332$\pm$0.001 & B95                          &  3.60 & M17a & -4.39 \\
HD 35850     & 0.553 & 0.468$\pm$0.029 & W04,G06,W07,C07,S09,I10      &  0.97 & W11  & -3.48 \\
HD 36395     & 1.474 & 1.763$\pm$0.364 & W04,W07,C07,I10              & 33.61 & W11  & -4.68 \\
HD 36705     & 0.830 & 1.486$\pm$0.020 & G06                          &  0.51 & W11  & -2.93 \\
HD 37216     & 0.764 & 0.379$\pm$0.011 & W04,W07,I10                  & 14.60 & W11  & -4.60 \\
HD 37394     & 0.840 & 0.453$\pm$0.001 & B95                          & 10.78 & M17a & -4.71 \\
HD 37572     & 0.845 & 0.804$\pm$0.052 & H96,G06,C07,S09              &  4.52 & W11  & -3.26 \\
HD 38949     & 0.566 & 0.261$\pm$0.012 & W04,W07,S09                  &  7.60 & W11  & -4.72 \\
HD 39587     & 0.594 & 0.325$\pm$0.001 & B95                          &  5.36 & M17a & -4.53 \\
HD 45081     & 0.963 & 2.952$\pm$0.020 & G06                          &  2.66 & W11  & -2.84 \\
HD 59747     & 0.863 & 0.536$\pm$0.034 & G03,W04,I10                  &  8.03 & W11  & -4.46 \\
\hline
\end{tabular}
\end{small}
\end{center}
\end{table*}
\setcounter{table}{0}
\begin{table*}
\caption{continued.}
\smallskip
\begin{center}
\begin{small}
\begin{tabular}{l c c l c c c l}
\hline\hline
\noalign{\smallskip}
Object & $B-V$ & S & Ref. & Per [day] & Ref & log L$_{\rm{X}}$/L$_{\rm{bol}}$ \\
\hline
\noalign{\smallskip}
HD 70516     & 0.652 & 0.396$\pm$0.035 & W04,W07,I10                  &  3.97 & W11  & -4.35 \\
HD 72905     & 0.618 & 0.367$\pm$0.001 & B95                          &  5.22 & M17a & -4.51 \\
HD 75302     & 0.689 & 0.261$\pm$0.003 & W04,W07,S09                  & 16.40 & W11  & -4.99 \\
HD 75393     & 0.536 & 0.369$\pm$0.037 & W04,W07,S09,I10              &  2.08 & W11  & -4.45 \\
HD 75732     & 0.869 & 0.165$\pm$0.008 & G03,W04,I10                  & 37.40 & M17a &       \\
HD 76218     & 0.771 & 0.375$\pm$0.007 & G03,W04,W07,I10              &  9.20 & W11  & -4.76 \\
HD 77407     & 0.609 & 0.386$\pm$0.029 & W04,W07,I10                  &  2.86 & W11  & -3.96 \\
HD 78366     & 0.585 & 0.248$\pm$0.001 & B95                          &  9.67 & W11  & -4.72 \\
HD 79211     & 1.420 & 1.909$\pm$0.065 & D91,W04,I10                  & 10.17 & W11  & -5.18 \\
HD 82558     & 0.933 & 1.458$\pm$0.110 & W04,W07                      &  1.70 & W11  & -3.51 \\
HD 82443     & 0.779 & 0.635$\pm$0.001 & B95                          &  5.41 & W11  & -3.83 \\
HD 82885     & 0.770 & 0.284$\pm$0.001 & B95                          & 18.60 & W11  & -5.01 \\
HD 85301     & 0.718 & 0.345$\pm$0.017 & W04,W07,I10                  &  7.47 & W11  & -5.06 \\
HD 87424     & 0.891 & 0.522$\pm$0.036 & W04,G06,I10,J11              & 10.74 & W11  & -4.98 \\
HD 88230     & 1.326 & 1.600$\pm$0.078 & D91,G03,I10                  & 11.67 & W11  & -5.23 \\
HD 90905     & 0.562 & 0.330$\pm$0.028 & W04,W07,I10                  &  2.60 & W11  & -4.20 \\
HD 92855     & 0.565 & 0.298$\pm$0.018 & W04,W07,I10                  &  4.41 & W11  & -4.51 \\
HD 95188     & 0.760 & 0.436$\pm$0.030 & W04,W07,I10                  &  6.81 & W11  & -4.52 \\
HD 95650     & 1.437 & 4.007$\pm$0.016 & W04,I10                      &  2.94 & W11  & -3.80 \\
HD 95735     & 1.502 & 0.424$\pm$0.001 & B95                          & 48.00 & W11  & -5.04 \\
HD 97334     & 0.600 & 0.335$\pm$0.001 & B95                          &  7.94 & M17a & -4.57 \\
HD 97658     & 0.845 & 0.176$\pm$0.011 & G03,W04,I10                  & 38.50 & H11  &       \\
HD 98712     & 1.340 & 2.892$\pm$0.435 & G06,S09                      & 11.60 & W11  & -4.16 \\
HD 100180    & 0.570 & 0.165$\pm$0.001 & B95                          & 14.00 & W11  & -5.94 \\
HD 101472    & 0.549 & 0.267$\pm$0.026 & W04,W07,I10                  &  4.46 & W11  & -4.61 \\
HD 101501    & 0.723 & 0.311$\pm$0.001 & B95                          & 16.06 & M17a & -5.15 \\
HD 103095    & 0.754 & 0.188$\pm$0.001 & B95                          & 31.10 & D92  &       \\
HD 104576    & 0.708 & 0.411$\pm$0.008 & W07,I10,J11                  &  9.10 & W11  & -3.91 \\
HD 106516    & 0.470 & 0.208$\pm$0.001 & B95                          &  6.91 & W11  & -5.82 \\
HD 113449    & 0.847 & 0.566$\pm$0.017 & G03,W07,J11,A11              &  6.47 & W11  & -4.16 \\
HD 114378    & 0.455 & 0.244$\pm$0.001 & B95                          &  3.00 & N84  &       \\
HD 114710    & 0.572 & 0.201$\pm$0.001 & B95                          & 12.35 & W11  & -5.66 \\
HD 115043    & 0.603 & 0.327$\pm$0.001 & B95                          &  5.87 & M17a & -4.63 \\
HD 115383    & 0.585 & 0.313$\pm$0.001 & B95                          &  3.33 & W11  & -4.41 \\
HD 115404    & 0.926 & 0.535$\pm$0.001 & B95                          & 18.47 & W11  & -5.08 \\
HD 115617    & 0.709 & 0.162$\pm$0.001 & B95                          & 29.00 & W11  & -6.90 \\
HD 118100    & 1.210 & 3.571$\pm$0.110 & G03,C07                      &  3.96 & W11  & -3.42 \\
HD 120136    & 0.508 & 0.191$\pm$0.001 & B95                          &  3.05 & M17b & -5.12 \\
HD 128311    & 0.973 & 0.624$\pm$0.038 & W04,G06,I10                  & 11.54 & W11  & -4.44 \\
HD 129333    & 0.626 & 0.544$\pm$0.001 & B95                          &  2.67 & W11  & -3.56 \\
HD 128621    & 0.900 & 0.191$\pm$0.011 & H96,G06,C07,S09              & 36.20 & W11  & -6.30 \\
HD 130307    & 0.893 & 0.389$\pm$0.030 & G03,W04,I10                  & 21.79 & W11  & -4.93 \\
HD 131156 A  & 0.720 & 0.461$\pm$0.001 & B95                          &  6.31 & W11  & -4.41 \\
HD 131156 B  & 1.156 & 1.381$\pm$0.001 & B95                          & 11.94 & W11  & -4.54 \\
HD 131977    & 1.024 & 0.523$\pm$0.033 & H96,C04,W04,G06,C07,S09,I10  & 32.50 & M17a &       \\
HD 132173    & 0.554 & 0.298$\pm$0.007 & H96,W04,W07,I10              &  4.92 & W11  & -4.61 \\
HD 133295    & 0.573 & 0.310$\pm$0.013 & H96,W04,G06,W07,I10          &  5.00 & W11  & -4.64 \\
HD 134319    & 0.677 & 0.419$\pm$0.032 & D91,W04,W07,I10              &  4.43 & W11  & -4.17 \\
HD 141004    & 0.604 & 0.155$\pm$0.001 & B95                          & 25.80 & W11  & -6.20 \\
HD 141272    & 0.801 & 0.439$\pm$0.018 & G03,W04,I10                  & 14.05 & W11  & -4.34 \\
HD 143761    & 0.612 & 0.150$\pm$0.001 & B95                          & 17.00 & B98  &       \\
HD 145229    & 0.604 & 0.282$\pm$0.017 & W04,W07,I10                  &  8.40 & W11  & -4.62 \\
HD 149661    & 0.827 & 0.339$\pm$0.001 & B95                          & 21.07 & W11  & -5.08 \\
HD 150554    & 0.591 & 0.189$\pm$0.007 & W04,W07,I10                  & 10.80 & W11  & -5.15 \\
HD 152391    & 0.749 & 0.393$\pm$0.001 & B95                          & 11.43 & W11  & -4.40 \\
HD 152555    & 0.591 & 0.359$\pm$0.020 & W04,W07,I10                  &  2.77 & W11  & -4.19 \\
HD 154417    & 0.578 & 0.269$\pm$0.001 & B95                          &  7.80 & W11  & -4.95 \\
HD 155885    & 0.855 & 0.384$\pm$0.001 & B95                          & 21.11 & W11  & -5.04 \\
HD 155886    & 0.855 & 0.375$\pm$0.001 & B95                          & 20.69 & W11  & -5.02 \\
HD 156026    & 1.144 & 0.770$\pm$0.001 & B95                          & 18.00 & W11  & -5.28 \\
HD 157881    & 1.359 & 1.684$\pm$0.109 & D91,G03,W04,I10              & 11.94 & W11  & -5.02 \\
HD 160346    & 0.959 & 0.300$\pm$0.001 & B95                          & 36.40 & W11  & -5.59 \\
HD 165341 A  & 0.860 & 0.392$\pm$0.001 & B95                          & 19.70 & W11  & -5.21 \\
\hline
\end{tabular}
\end{small}
\end{center}
\end{table*}
\setcounter{table}{0}
\begin{table*}
\caption{continued.}
\smallskip
\begin{center}
\begin{small}
\begin{tabular}{l c c l c c c l}
\hline\hline
\noalign{\smallskip}
Object & $B-V$ & S & Ref. & Per [day] & Ref & log L$_{\rm{X}}$/L$_{\rm{bol}}$ \\
\hline
\noalign{\smallskip}
HD 166435    & 0.633 & 0.432$\pm$0.024 & G03,W04,W07,I10              &  3.80 & W11  & -4.14 \\
HD 166620    & 0.876 & 0.190$\pm$0.001 & B95                          & 42.40 & W11  & -6.18 \\
HD 168603    & 0.771 & 0.376$\pm$0.003 & W04,I10                      &  4.83 & W11  & -4.52 \\
HD 170778    & 0.619 & 0.308$\pm$0.009 & W04,W07,I10                  &  6.46 & W11  & -4.19 \\
HD 172649    & 0.525 & 0.302$\pm$0.020 & W04,W07,I10                  &  3.94 & W11  & -4.74 \\
HD 178428    & 0.705 & 0.154$\pm$0.001 & B95                          & 22.00 & W11  & -5.51 \\
HD 180161    & 0.804 & 0.385$\pm$0.013 & W04,I10                      &  5.49 & W11  & -4.61 \\
HD 182101    & 0.458 & 0.216$\pm$0.001 & B95                          &  5.02 & R87  &       \\
HD 186427    & 0.661 & 0.152$\pm$0.003 & D91,G03,W04,G06,I10          & 31.00 & HOO  &       \\
HD 187691    & 0.563 & 0.149$\pm$0.001 & B95                          & 15.00 & R87  &       \\
HD 187897    & 0.647 & 0.257$\pm$0.018 & W04,W07,S09,I10              & 11.00 & W11  & -5.01 \\
HD 189733    & 0.932 & 0.498$\pm$0.020 & G03,W04,I10                  & 11.80 & W11  & -4.66 \\
HD 190007    & 1.128 & 0.746$\pm$0.001 & B95                          & 28.95 & W11  & -5.11 \\
HD 190406    & 0.600 & 0.194$\pm$0.001 & B95                          & 13.94 & W11  & -5.68 \\
HD 193017    & 0.567 & 0.227$\pm$0.009 & W04,W07,S09,I10              &  8.90 & W11  & -5.25 \\
HD 199019    & 0.767 & 0.465$\pm$0.027 & W04,W07,I10                  &  6.54 & W11  & -4.42 \\
HD 200746    & 0.654 & 0.321$\pm$0.005 & W04,W07                      &  7.61 & W11  & -4.67 \\
HD 201091    & 1.069 & 0.658$\pm$0.001 & B95                          & 35.37 & W11  & -5.39 \\
HD 201092    & 1.309 & 0.986$\pm$0.001 & B95                          & 37.84 & W11  & -5.58 \\
HD 201219    & 0.692 & 0.281$\pm$0.007 & W04,W07,I10                  & 16.00 & W11  & -4.96 \\
HD 201989    & 0.689 & 0.317$\pm$0.017 & H96,W04,G06,W07,I10          & 14.80 & W11  & -4.63 \\
HD 203030    & 0.750 & 0.451$\pm$0.036 & W04,W07,I10                  &  6.67 & W11  & -4.37 \\
HD 204277    & 0.529 & 0.261$\pm$0.021 & W04,W07,I10                  &  4.50 & W11  & -4.61 \\
HD 205905    & 0.623 & 0.256$\pm$0.010 & H96,W04,G06,W07,I10          & 11.18 & W11  & -5.14 \\
HD 206374    & 0.686 & 0.247$\pm$0.032 & G03,W04,W07,I10              & 19.20 & W11  & -5.00 \\
HD 206860    & 0.587 & 0.330$\pm$0.001 & B95                          &  4.86 & W11  & -4.39 \\
HD 209253    & 0.504 & 0.305$\pm$0.018 & W04,G06,W07,I10              &  3.00 & W11  & -4.50 \\
HD 209393    & 0.693 & 0.358$\pm$0.013 & W04,W07,I10                  &  7.20 & W11  & -4.87 \\
HD 209779    & 0.674 & 0.306$\pm$0.026 & W04,G06,W07                  & 10.10 & W11  & -4.40 \\
HD 210667    & 0.812 & 0.326$\pm$0.017 & D91,G03,W04,W07,I10          &  9.08 & W11  & -4.82 \\
HD 216803    & 1.094 & 1.068$\pm$0.072 & H96,W04,C07,I10              &  9.87 & W11  & -4.57 \\
HD 217014    & 0.666 & 0.149$\pm$0.001 & B95                          & 37.00 & W11  & -7.20 \\
HD 220182    & 0.801 & 0.483$\pm$0.046 & D91,G03,W04,I10              &  7.49 & W11  & -4.41 \\
HD 245409    & 1.415 & 3.243$\pm$0.341 & D91,W04,I10                  & 12.29 & W11  & -4.13 \\
HD 284253    & 0.813 & 0.412$\pm$0.016 & D91,I10                      & 10.26 & W11  & -4.73 \\
HD 284414    & 0.908 & 0.430$\pm$0.073 & D91,I10                      &  9.90 & W11  & -4.65 \\
HD 285690    & 0.980 & 0.251$\pm$0.020 & D91                          & 12.64 & W11  & -4.45 \\
HD 285773    & 0.831 & 0.417$\pm$0.017 & D91,W07,I10                  & 11.38 & W11  & -4.65 \\
HD 285968    & 1.523 & 1.400$\pm$0.077 & D91,W04,I10                  & 38.92 & W11  & -4.79 \\
HIP 6276     & 0.791 & 0.547$\pm$0.024 & W04,G06,W07,I10              &  6.40 & W11  & -4.30 \\
HIP 49986    & 1.487 & 1.964$\pm$0.067 & W04,I10                      & 21.56 & W11  & -4.79 \\
HIP 63510    & 1.448 & 9.234$\pm$0.041 & D91,I10                      &  1.54 & W11  & -3.36 \\
HIP 86087    & 1.456 & 2.269$\pm$0.019 & D91                          & 18.60 & W11  & -4.97 \\
\hline
\end{tabular}
\tablefoot{The $B - V$ values are taken from from \cite{HIPPARCOS1997ESA}, except for the Sun. The solar  
$B-V$-value is taken from \citet{strobel1996}. The log L$_{\rm{X}}$/L$_{\rm{bol}}$ values are taken from \citet{wright2011ApJ.743.48W}.}
\tablebib{(B95)~\citet{b95},(W11)~\citet{wright2011ApJ.743.48W},(W04)~\citet{Wright2004yCat.21520261W},
  (I10)~\citet{Isaacson2010ApJ.725.875I}, (H96)~\citet{Henry1996AJ.111.439H},(G03)~\citet{Gray2003AJ.126.2048G},(G06)~\citet{Gray2006AJ.132.161G},(D91)~\citet{Duncan2005yCat.3159.0D},(W07)~\citet{White2007AJ.133.2524W},(S09)~\citet{Schroeder2008yCat.34931099S},(C04)~\citet{Cincunegui2003yCat.34140699C},(H00)~\citet{henry2000},(R87)~\citet{rutten1987},(B98)~\citet{b98},(M17a)~\citet{Mittag2017A&A...607A..87M},(M17b)~\citet{Mittag2017A&A...600A.119M},(D92)~\citet{dobson1992},(N84)~\citet{noyes1984},(H14)~\citet{Howard2014ApJ...794...51H}}
\end{small}
\end{center}
\end{table*}

\end{appendix}
\end{document}